\documentclass[%
reprint,
superscriptaddress,
amsmath,amssymb,
aps,
floatfix,
]{revtex4-1}

\usepackage[english]{babel}
\usepackage[dvips]{graphicx}
\usepackage{color}
\usepackage{latexsym}
\usepackage{amsmath}
\usepackage{amsthm}
\usepackage{amsfonts}
\usepackage{amssymb}
\usepackage{romannum}
\usepackage[T1]{fontenc}
\usepackage{braket}
\usepackage{dsfont}
\usepackage{xcolor}
\usepackage{bm}

\begin{document}
	
\pagenumbering{arabic}
	
\title{Second Chern Number and Non-Abelian Berry Phase  in Topological Superconducting Systems}
\author{H. Weisbrich}
\affiliation{Fachbereich Physik, Universit{\"a}t Konstanz, D-78457 Konstanz, Germany}
\author{R. L. Klees}
\affiliation{Fachbereich Physik, Universit{\"a}t Konstanz, D-78457 Konstanz, Germany}
\author{G. Rastelli}
\affiliation{Fachbereich Physik, Universit{\"a}t Konstanz, D-78457 Konstanz, Germany}
\affiliation{Zukunftskolleg, Universit{\"a}t Konstanz, D-78457 Konstanz, Germany}
\author{W. Belzig}
\email{Corresponding author: wolfgang.belzig@uni-konstanz.de}
\affiliation{Fachbereich Physik, Universit{\"a}t Konstanz, D-78457 Konstanz, Germany}

\begin{abstract}
Topology ultimately unveils the roots of the perfect quantization
observed in complex systems. The 2D quantum Hall effect is the
celebrated archetype. Remarkably, topology can manifest itself even in higher-dimensional spaces in which control parameters play the role of extra, synthetic dimensions.
However, so far, a very limited number of implementations of higher-dimensional topological systems have been proposed, a notable example being the so-called 4D quantum Hall effect.
Here we show that mesoscopic superconducting systems can implement higher-dimensional topology and represent a formidable platform to study a quantum 
system with a purely nontrivial second Chern number. 
We demonstrate that the integrated absorption intensity in designed microwave spectroscopy is quantized and the integer is directly related to the second Chern number. Finally, we show  that these systems also admit a non-Abelian Berry phase. Hence,  they also realize an enlightening paradigm of  topological non-Abelian systems in higher dimensions. 
\end{abstract}

\date{\today}

\maketitle

%
%

\section{Introduction}
Topology ultimately explains the origin of phenomena that at first glance appear extremely fragile.
The standard example is a gas of electrons confined in a plane under the effect of a strong magnetic field (the so-called 2D quantum Hall effect)  \cite{klitzing1980new,TKKN}, in which the conductance results to be an integer in units of fundamental constants in physics, the elementary electron charge and the Planck constant. 
Real 2D electron systems are very far from being close to the simplified theoretical models and many microscopic details are inaccessible and unknown, such as structural disorder, atomic impurities, or edge configurations. 
Why simple toy models can explain such perfect quantization in real and complex systems remained mysterious until topology disclosed 
that all these systems are equivalent in the topological sense.
Indeed, for the 2D quantum Hall effect, the quantized conductance is related to the first Chern number of the occupied electronic bands \cite{lhatsugai1993chern}. 

More generally, the concept of topology brought a unique understanding of the physics of materials, such as for topological insulators which are insulating in the bulk, yet have conducting surface/edge states \cite{Hasan:2010ku}. 
Other examples are superconducting (SC) junctions made of nanowires that have opened the route to the engineering of topological superconductors \cite{Sato:2017go}.
Such SC systems can host exotic states, the so-called Majorana zero-energy modes, which attracted a broad interest  fueled by the promise of topologically protected quantum computing \cite{sarma2015majorana,deng2016majorana,aasen2016milestones,karzig2017scalable}.
More specifically, Majorana-based qubits illustrate the idea of holonomic quantum computation  in a paradigmatic way which is based on the concept of a generalized non-Abelian Berry phase in a degenerate ground-state subspace \cite{wilczek1984app}.
In this scheme, after an adiabatic and cyclic change of the system's parameters, the initial state in the ground state subspace can be transformed  
to a linear combination of the subspace basis which depends on the evolution of the parameters.
Thus, adiabatic cyclic paths implement arbitrary unitary transformations {\textemdash} or quantum gates in the language of quantum information {\textemdash} in the degenerate subspace \cite{pachos1999non,zanardi1999holonomic} whose speed is limited by the inverse of the energy difference with the first excited state (or energy gap) \cite{albash2018adiabatic}.

The continuous search for new types of topological quantum matter has recently led to the discovery of topologically nontrivial quantum states in conventional multiterminal Josephson junctions \cite{riwar2016multi,eriksson2017topological,xie2017topological,meyer2017nontrivial,xie2018weyl,deb2018josephson,xie2019topological,klees2020microwave}. 
Due to their scalability, SC Josephson circuits have already opened the path towards realistic implementations of quantum technologies \cite{blais2020quantum,martinis2020quantum}. 
These systems are based on the Josephson effect \cite{josephson1962possible,Golubov2004JJ} in which a supercurrent flows through a weak link which can be a tunnel junction, a molecule, or a quantum dot between two superconductors \cite{glazman:89,buitelaar2002quantum,van2006supercurrent,de2010hybrid}.
Microscopically, this supercurrent is carried by the so-called Andreev bound states which are localized at the weak link \cite{pillet:10}.  
Because they have discrete energies inside the SC gap, Andreev bound states can be coherently manipulated and experimentally accessed by microwave spectroscopy \cite{bretheau2013exciting,janvier2015coherent,van2017microwave,tosi2019spin} and supercurrent spectroscopy \cite{bretheau2013supercurrent}. 
Andreev bound states can be exploited to encode information in novel types of SC qubits which are not based on collective electromagnetic degrees of freedom, such as charge, SC phase, or flux, but on microscopic quasiparticle states inherent to SC weak links \cite{zazunov:2003,chtchelkatchev:2003}.
At the same time, such systems, as for example multiterminal SC Josephson junctions, have turned out to be an ideal platform where synthetic topological materials can be engineered almost at will \cite{riwar2016multi,eriksson2017topological,xie2017topological,meyer2017nontrivial,xie2018weyl,deb2018josephson,klees2020microwave,xie2019topological}.

Interestingly, topology is not restricted to low-dimensional systems as the canonical example of the 2D quantum Hall effect, but it can also emerge 
in higher-dimensional spaces in which controlling parameters play the role of extra synthetic dimensions. 
The intriguing case is if a system may be topologically trivial within a restricted (2D) subspace (namely it has a vanishing first Chern number) yet can show nontrivial topology in higher dimensions with nonzero higher-dimensional invariants (e.g., a nonzero second Chern number in 4D space). 
Indeed, the concept of the 2D quantum Hall effect has been extended to the 4D quantum Hall effect with the quantized nonlinear Hall response determined by the second Chern number \cite{zhang2001four}.
Even though this situation does not naturally arise in solid state systems due to limited dimensionality, there are several possibilities to create the 4D space artificially. 
It has been theoretically proposed to implement the 4D quantum Hall effect using the internal transitions of cold atoms in a 3D optical lattice \cite{price2015four} or in a 2D crystal with modulated on-site potentials \cite{kraus2013four}.
The latter proposal has been experimentally realized in a system of an angled optical superlattice of ultracold bosonic atoms \cite{lohse2018exploring} and in tunable 2D arrays of photonic waveguides \cite{zilberberg2018photonic}.
Other theoretical proposals to explore topological higher-dimensional systems are based on the simulation of a non-Abelian Yang monopole by cyclically coupling the hyperfine structure of Rubidium \cite{sugawa2018second} 
and on the implementation of one-way optical waveguides with designed spatial modulations \cite{lu2018topological}. 
Finally, even topologically protected Majorana modes can arise from a nontrivial second Chern number \cite{chan2017non}. 
 
Overall, so far, a very limited number of implementations of higher-dimensional topological systems have been theoretically proposed and only a 
few of them have been experimentally realized.
In this work, we show that nanoscale SC systems can implement higher-dimensional topology and represent a formidable  platform to study a quantum system with a nontrivial second Chern number. 
In our proposal, the order is not limited by spatial dimensions since the additional dimensions can be easily implemented by increasing the number of independent SC phases applied at different SC contacts.
We propose systems formed by multiterminal SC contacts embedding quantum dots and that are characterized by a nontrivial second Chern number. 
In the topological regime, for large energy gap and perturbative tunneling coupling between the SC leads, the system possesses one pair of twofold degenerate states with a non-Abelian Berry phase.
We show that the nontrivial 4D topology manifests itself in the integrated microwave response of the system by means of nonadiabatic effects \cite{kolodrubetz2016measuring}.
We present a method to measure the non-Abelian Berry curvature in the degenerate subspace using a suitable measurement protocol.

First, we discuss a deterministic scheme for the initial preparation of a target state in the ground-state subspace.
The scheme is based on the adiabatic sweep from the nondegenerate case to the degenerate case followed by specific adiabatic cyclic paths (non-Abelian Berry rotations) of the SC phases.
Since microwave spectroscopy on discrete Andreev bound states is a well established technique by now \cite{bretheau2013exciting,janvier2015coherent,van2017microwave,tosi2019spin}, as a second step we will apply designed polarized microwave spectroscopy \cite{klees2020microwave} to access the diagonal and off-diagonal elements of the non-Abelian Berry curvature in the degenerate subspace.
Finally, the difference of the oscillator strengths for different circular polarizations integrated over the 4D parameter space corresponds to the integration of the local Berry curvature and, therefore, the result will be directly related to the second Chern number.
%
%
%
%
%
%
%
%
%
%
%
\begin{figure*}
	\includegraphics[width=0.8\textwidth]{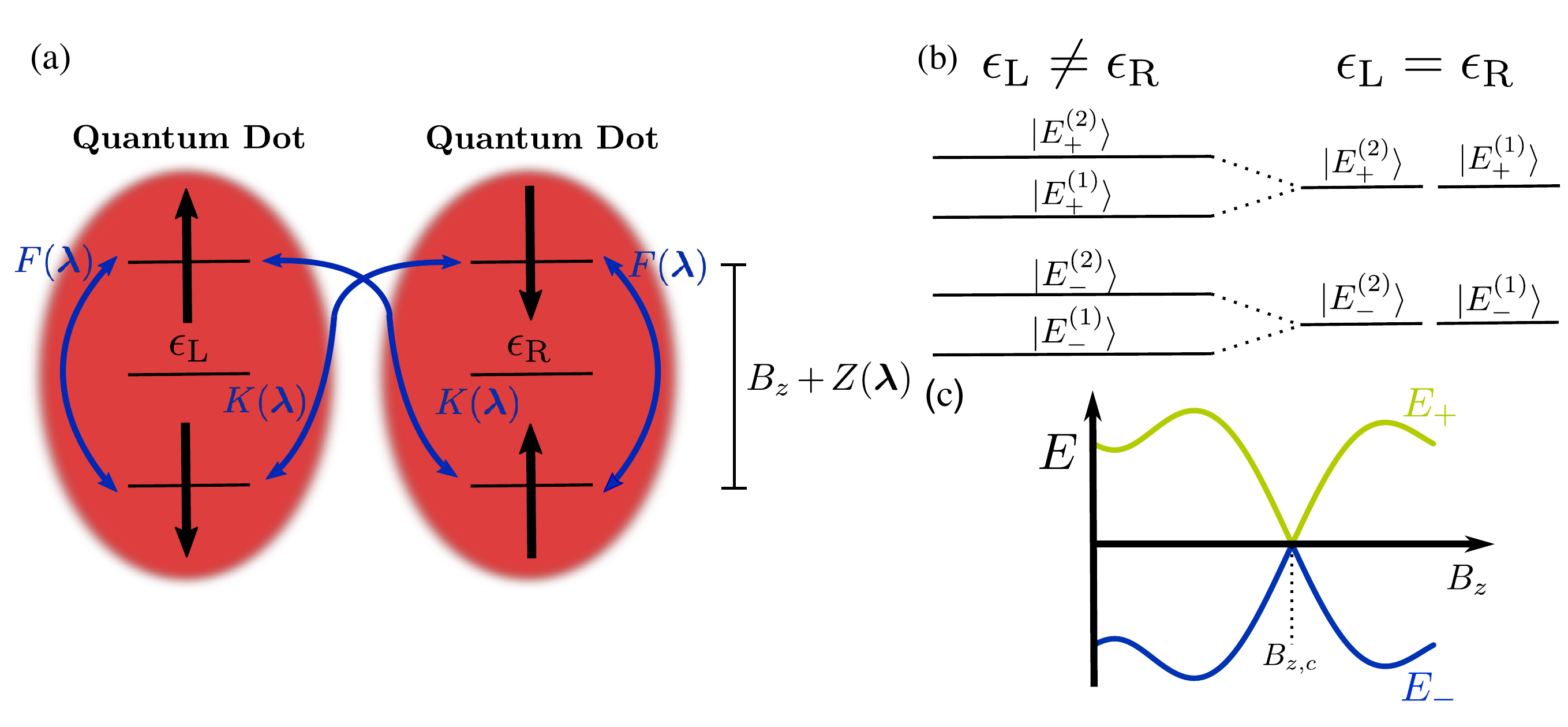}
	\caption{(a) Double-quantum-dot system with mediated interactions $F(\bm{\lambda})$, $K(\bm{\lambda})$, and $Z(\bm{\lambda})$ via four parameters $\bm{\lambda} = (\lambda_1, \lambda_2, \lambda_3 , \lambda_4)$. (b) For $\epsilon_\mathrm{L} \neq \epsilon_\mathrm{R}$, there are two pairs of states in the double-dot system. At the symmetric point for $\epsilon_\mathrm{L} = \epsilon_\mathrm{R}$, there is only one pair of twofold degenerate states. (c) Sketch of an energy crossing of the two twofold degenerate bands appearing at the crossing point $B_{z,c}$ for $\bm{\lambda}=\bm{\lambda}_c$, which leads to a change of the second Chern number.}
	\label{fig:Figure1}
\end{figure*}
\section{Model and results} 
\label{sec:results}
\subsection{Model system consisting of two quantum dots}
\label{sec:modelAndEffectiveHamiltonianOfTheDot}
Our starting point are two quantum dots (left L and right R) with two spin levels whose 
average  energies are $\epsilon_\mathrm{L}$ and $\epsilon_\mathrm{R}$, 
respectively. 
A local Zeeman field $B_z$ is externally applied with opposing direction on each dot. 
%
%
Such a simple double-dot system is coupled to a network composed by different SC contacts which we call hereafter environment.
For the sake of simplicity, we assume that all the SC contacts are similar, namely, with the same SC gap $\Delta$ and at the same chemical potential set to zero as a reference.
The SC phases associated with the environment provide the synthetic gauge fields composed by four independent parameters 
$\bm{\lambda} = (\lambda_1 \, , \lambda_2 \, , \lambda_3 \, , \lambda_4)$ from which we define topological properties.
The independent phases are used to implement a four-dimensional space providing nontrivial topology in terms of higher-dimensional invariants.
Indeed, after integrating out the environment, the tunnel coupling of the dots with the SC environment results in an effective phase-dependent Hamiltonian acting solely on the double-dot system which defines the energy spectrum of the lowest discrete states inside the SC gap.
Naturally, the final result depends on the structure of the environment and what kind of interaction or structure we synthetize in the double-dot system. 
The advantage of this general method is the large freedom in constructing a specific effective Hamiltonian of the double-dot system providing all the necessary extra dimensions. 

\begin{figure*}
	\includegraphics[width=0.9\textwidth]{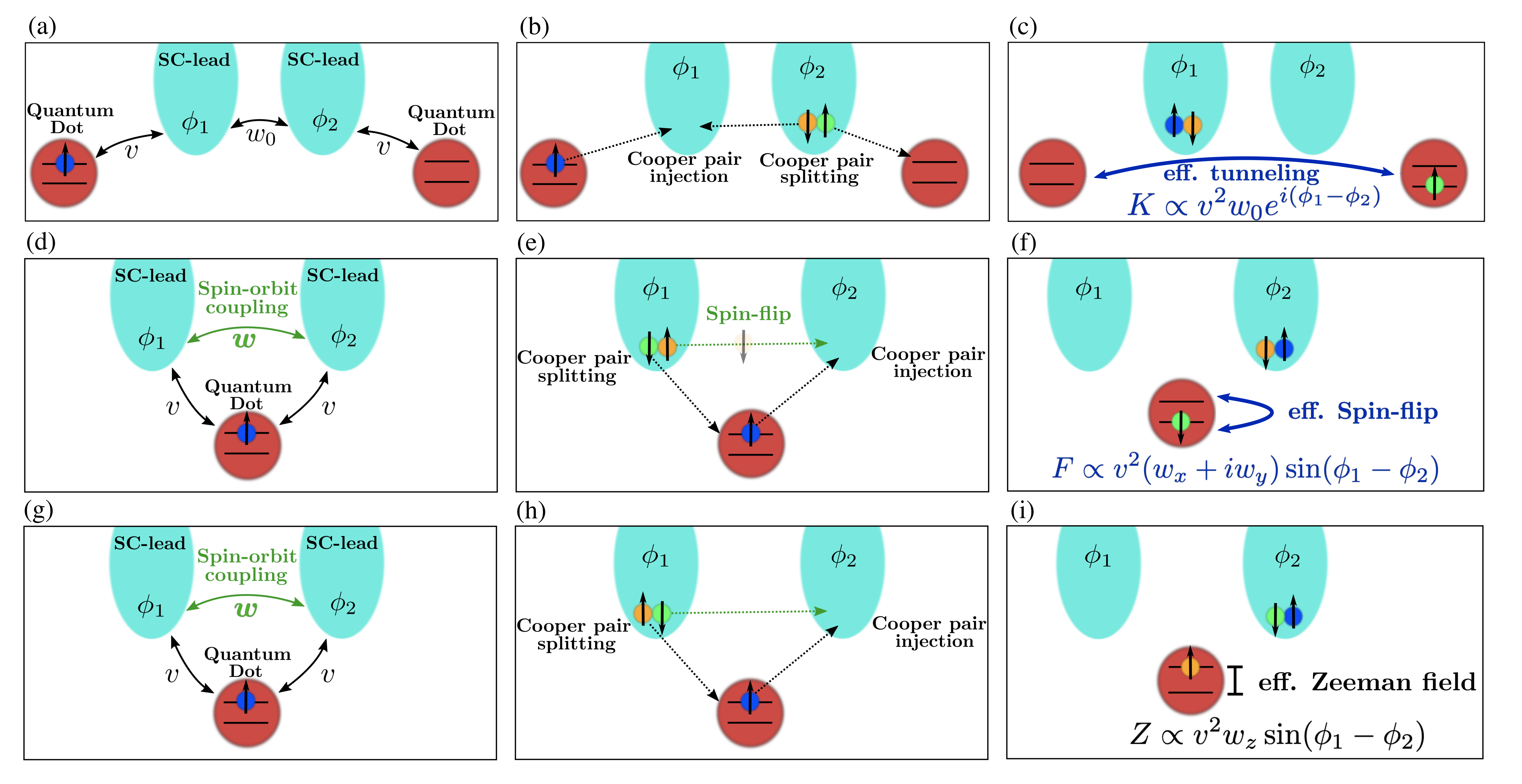}
	\caption{
		(a)  Two dots (red) coupled to two SC leads with $w_0\ll v$. (b) The effective hopping between the dots is mediated by two crossed Andreev reflection processes in which a Cooper pair is combined (left lead) and one Cooper pair is splitted (right lead). (c) This mechanism leads to an effective  tunneling $K\propto v^2w_0 e^{i(\phi_1-\phi_2)}$ depending on the two phases of the leads. 
		(d) One dot (red) coupled to two SC leads which are itself connected via spin-orbit coupling $\bm{w}$. (e) The spin-flip of an electron on the dot is mediated by two crossed Andreev reflection processes in which a Cooper pair is combined (right lead) and one Cooper pair is splitted (left lead). For this mechanism, a Rashba-like spin-orbit interaction between the superconductors is needed.  (f) The process results in an effective phase dependent spin-flip term $F\propto  v^2(w_x+iw_y)\sin(\phi_1-\phi_2)$ on the dot.	(g) One dot (red) coupled to two SC leads which are itself coupled via spin-orbit interaction $\bm{w}$. (h) The mechanism is mediated by two crossed Andreev reflection processes in which a Cooper pair is combined (right lead) and one Cooper pair is splitted (left lead). (i) The process results in an effective Zeeman term $Z \propto v^2w_z\sin(\phi_1-\phi_2)$ on the dot.}
	\label{fig:Figure2}
\end{figure*}

As depicted in Fig.~\ref{fig:Figure1}(a), the aimed and minimal effective Hamiltonian of the double-dot system showing a nontrivial second Chern number has the form
%
%
%
%
\begin{align}
H &= 
\sum_{\alpha= \mathrm{L,R}}
\sum_{\sigma=\uparrow, \downarrow} 
 \epsilon_\alpha \, d_{\alpha\sigma}^\dagger d_{\alpha\sigma}^{\phantom\dag}  
\nonumber\\&\quad +
\bigl[ B_{z} + Z(\bm{\lambda}) \bigr]
\sum_{\sigma=\uparrow, \downarrow}  
\sigma (d_{\mathrm{L}\sigma}^\dagger d_{\mathrm{L}\sigma}^{\phantom\dag}   - d_{\mathrm{R}\sigma}^\dagger d_{\mathrm{R}\sigma}^{\phantom\dag}   )
\nonumber \\
&\quad 
+ 
\sum_{\sigma=\uparrow, \downarrow} 
 \bigl[ K(\bm{\lambda}) \, d_{\mathrm{L}\sigma}^\dagger d_{\mathrm{R}\sigma}^{\phantom \dag} + \mathrm{h.c.} \bigr]
\nonumber\\&\quad +
\bigl[
F(\bm{\lambda}) \, (d_{\mathrm{L}\uparrow}^\dagger d_{\mathrm{L}\downarrow}^{\phantom\dag} -d_{\mathrm{R}\uparrow}^\dagger d_{\mathrm{R}\downarrow}^{\phantom\dag}) + \mathrm{h.c.} \bigr] ,
\label{Hamstructure}
\end{align}
%
%
%
%
%
where $d_{\alpha\sigma}^\dagger$  $(d_{\alpha\sigma})$  is the creation (annihilation) operator of an electron with spin $\sigma = \uparrow,\downarrow$ on the left ($\alpha=\mathrm{L}$) or the right ($\alpha=\mathrm{R}$) dot, respectively, and h.c. denotes the Hermitian conjugate.
As anticipated, the effective Hamiltonian possesses  an additional parameter-dependent Zeeman term $Z(\bm{\lambda})$ on both dots with reversed direction. 
As shown in Fig.~\ref{fig:Figure1}(b), such a Hamiltonian provides a twofold degenerate ground state $\ket{E_{-}^{(1/2)}}$ and a twofold degenerate excited state $\ket{E_{+}^{(1/2)}}$ at the symmetric point $\epsilon_\mathrm{L} = \epsilon_\mathrm{R} = \epsilon_0$ with energies $E_{\pm} = \epsilon_0 \pm\sqrt{|B_z+Z(\bm{\lambda})|^2+|K(\bm{\lambda})|^2+|F(\bm{\lambda})|^2}$.
To engineer this Hamiltonian, we need a parameter-dependent coupling between both dots via the complex hopping $K(\bm{\lambda})$. 
Finally, we also demand a complex parameter-dependent spin-flip term $F(\bm{\lambda})$ on each dot with reversed sign between the left and the right dot. 
We show how to engineer these terms using the SC multiterminal environment in Fig.~\ref{fig:Figure2}.
For instance, as depicted in Figs.~\ref{fig:Figure2}a-c, an effective phase-dependent hopping $K \propto e^{i (\phi_2-\phi_1)}$ between the two dots can be realized via two crossed Andreev reflection (CAR) processes in which Cooper pairs break in one lead and recombine in another lead, effectively transferring an electron from one to the other dot.
Similarly, as shown in Figs.~\ref{fig:Figure2}d-f, one can also engineer a phase-dependent effective local spin-flip $F\propto \sin(\phi_2-\phi_1)$ with the help of two spin-orbit coupled leads where again two CAR processes are involved.
Finally, an effective phase-dependent local Zeeman splitting $Z \propto \sin(\phi_1-\phi_2)$ on the dots is realized via a spin-dependent hopping between two SC leads that are coupled to the same dot, see Figs.~\ref{fig:Figure2}g-i .

The Hamiltonian in Eq.~\eqref{Hamstructure} conserves the fermionic parity.
Assuming strong Coulomb interactions on the dots, local Andreev reflection, in which a single Cooper pair is injected from an SC lead to a single dot, is forbidden.
This allows us to analyze the Hamiltonian in Eq.~\eqref{Hamstructure} by focusing only on the odd-parity regime in which the full double-dot system is occupied by only a single electron.
%
%
In the following, as a proof of concept, we give two examples for the SC environments which allow for the presented structure and lead to a nontrivial second Chern number.
Although this scheme is totally general, we present two specific examples to illustrate the underlying ideas.
%
%
%
%
%
%
%
%

\subsubsection{Example A}
A first example of a topological higher-dimensional system is formed by a multiterminal structure composed by nine standard BCS SC contacts described by four independent phases. 
Whereas gauge invariance allows us to set one phase to zero ($\phi_0=0$) as a reference, we set the other phases to values such that only four SC phases are independent, see Fig.~\ref{fig:Figure3}(a).
Hence, the parameter space is defined by these four phases, i.e.,  $\bm{\lambda}_\mathrm{A} :=  (\phi_1,\phi_2,\phi_3,\phi_4)$.
As shown in Fig.~\ref{fig:Figure3}(a), some contacts are tunnel-coupled to the left dot or the right dot and some contacts are also tunnel-coupled between each other.
In particular, the SC leads are coupled either via normal tunnel hoppings, described by a scalar parameter $w_0$ taking the form $w_0(\sigma_0\otimes \tau_3)$ in spin (Pauli matrices $\sigma_0, \ldots , \sigma_3)$ and Nambu (Pauli matrices $\tau_0 , \ldots , \tau_3$) space, respectively.
Or they are coupled via spin-orbit  tunnel coupling, described by the vectors $\bm{w}_j = (w_{x,j},w_{y,j},w_{z,j})^\mathrm{T}$, namely spin-flip processes 
are possible during the tunneling.
Here, the tunneling can be expressed as $i\bm{w}_j \cdot \bm{\sigma} \otimes \tau_3$ in spin-Nambu space (see Supplemental Material \cite{supplement}).
Assuming large local Coulomb interactions on the dots, Cooper pair injections are excluded and, in the odd-parity subspace, the resulting low-energy Hamiltonian for the lowest pair of states takes the form of Eq.~\eqref{Hamstructure} to first order in the coupling between the leads $(w_0,|\bm{w}_{j}|\ll v)$. 
Here, $v$ is the normal coupling of the dots to the environmental leads.
The explicit form of the terms reads \cite{supplement}
%
%
%
%
%
%
%
%
\begin{subequations}
	\begin{align}
		Z(\bm{\lambda}_\mathrm{A})&= - 2\epsilon_\mathrm{T} R_{z}(\bm{\lambda}_\mathrm{A}), 
		\\
		K(\bm{\lambda}_\mathrm{A})& =  v_0 - \epsilon_\mathrm{T} (e^{i\phi_1} + e^{-i\phi_2} ),
		\\
		F(\bm{\lambda}_\mathrm{A})& =-2\epsilon_\mathrm{T} \bigl( R_{x}(\bm{\lambda}_\mathrm{A}) - i R_{y}(\bm{\lambda}_\mathrm{A}) \bigr)  ,
	\end{align}
\end{subequations}
%
%
%
%
%
%
%
%
%
%
where $v_0$ is the direct coupling between the dots, and $\epsilon_\mathrm{T} = \pi^2 N_0^2 v^2w_0$ is the characteristic energy of the mediated hopping by the superconductors. In addition, $N_0$ is the normal density of states at the Fermi energy of the superconductors. 
The dimensionless vector $\bm{R}(\bm{\lambda}_\mathrm{A}) = [ \bm{w}_{1}\sin(\phi_3) + \bm{w}_{2} \sin(\phi_4-\phi_3) + \bm{w}_{3} \sin(\phi_4-\phi_2)] / w_0$ is 
associated with the spin-orbit tunnel hopping.
A more detailed discussion on the range of parameters and the stability of the topological phase is presented in the Supplemental Material \cite{supplement}. 
\\\\\\\\
\subsubsection{Example B}
\label{subsubsec:ExampleB}
\begin{figure*}
	\includegraphics[width=1\textwidth]{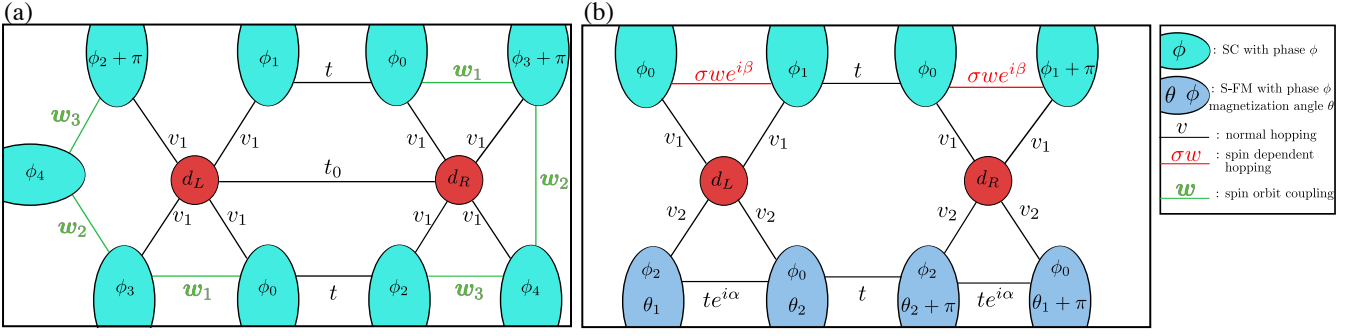}
	\caption{
		(a) Full system of Example A with an environmental network of nine SC leads coupled via normal hoppings ($w_0,v_0,v$) or via spin-orbit couplings ($\bm{w}_j$). Each lead has a SC phase $\phi_j$ and the dots have a magnetic field $B_z$ in opposing direction.
		(b) Full system of Example B with an environmental network of four SC leads with normal hopping ($v,w_0$) and spin dependent hopping ($\sigma w$), as well as four SC leads with a local exchange field $h$ with magnetization direction $\theta_j$ (with respect to the direction of magnetic field $B_z$ on the dots) coupled via normal hopping ($\tilde{v},\tilde{w}_0$). Each lead has a SC phase $\phi_j$ and the dots have a magnetic field $B_z$ in opposing direction.}
	\label{fig:Figure3}
\end{figure*}

As a second example, we consider an environment with eight leads, as depicted in Fig.~\ref{fig:Figure3}(b).
Four leads will be standard BCS superconductors described by the two SC phases $\phi_1$ and $\phi_0$ with $\phi_0=0$ as our choice of gauge.
These leads are coupled either by a spin-dependent hopping ($\sigma w_0e^{i\pi/2}$) with a complex phase or by a simple tunnel coupling ($w$), as illustrated in  Fig.~\ref{fig:Figure3}(b).
The other four leads are assumed to be superconductor-ferromagnet (S-FM) hybrid bilayers \cite{buzdin2005proximity}.
The latter contacts are described by the two SC phases  $\phi_2$ and  $\phi_0=0$. 
In addition, the local magnetizations are assumed to have the same magnitude $h$ but are tilted by relative angles $\theta_j$ with respect to the quantization axis associated with the Zeeman field $B_z$ in the dots.
%
%
Therefore, the parameter space for Example B is given by 
$\bm{\lambda}_\mathrm{B} :=  (\phi_1,\phi_2,\theta_1,\theta_2)$.
The superconductor-ferromagnet hybrid bilayer contacts are coupled via normal hoppings
($\tilde{w}_0$) . 
Again, in the first order of the coupling between the leads ($w_0\ll v$, $\tilde{w}_0\ll\tilde{v}$) and in the odd-parity regime (large local Coulomb interactions), the structure of the double-dot system takes the form of Eq.~\eqref{Hamstructure} with the following terms \cite{supplement}
%
%
%
%
%
%
%
%
%
%
%
%
%
%
%
%
%
%
%
%
\begin{subequations}
	\begin{align}
		Z(\bm{\lambda}_\mathrm{B})&=\frac{h \, \epsilon_\mathrm{h} }{\sqrt{\Delta^2-h^2}}\left(\cos(\theta_1)+\cos(\theta_2)\right)-2\epsilon_\mathrm{T}\sin(\phi_1)
		\\
		K(\bm{\lambda}_\mathrm{B}) &= -\Bigl( 
		\epsilon_\mathrm{T} e^{i\phi_1}
		+
		\frac{\tilde{\epsilon}_\mathrm{T}  (h^2 + \Delta^2 \, e^{-i\phi_2} )}{\Delta^2-h^2} 
		\Bigr)
		\\
		F(\bm{\lambda}_\mathrm{B})&=\frac{h \,  \epsilon_\mathrm{h}}{\sqrt{\Delta^2-h^2}}\big(\sin(\theta_1)+\sin(\theta_2)\big)
		\nonumber\\&\quad+
		i \frac{2h^2}{\Delta^2-h^2}\tilde{\epsilon}_\mathrm{T}\sin(\theta_2-\theta_1)
		\, ,
	\end{align}
\end{subequations}
%
%
%
%
%
%
%
%
%
%
as derived in the Supplemental Material \cite{supplement} with $\tilde{\epsilon}_\mathrm{T}=\pi^2 N_0^2 \tilde{v}^2 \tilde{w}_0$ and $\epsilon_\mathrm{h} = \pi N_0 \tilde{v}^2$, whereas $\Delta$ is the gap of the superconductors. 
The energy of both dots are renormalized by $\tilde{\epsilon}_0 = \epsilon_0 - h^2\tilde{\epsilon}_T  \sin(\phi_2) / (\Delta^2-h^2)$.
However, this renormalization does not influence our topological findings since the discrete low-energy spectrum of the system is still given by the eigenvalues 
$E_{\pm} = \tilde{\epsilon}_0 \pm \sqrt{|B_z + Z(\bm{\lambda}_\mathrm{B})|^2+|K(\bm{\lambda}_\mathrm{B})|^2+|F(\bm{\lambda}_\mathrm{B})|^2}$. 
A more detailed discussion on the range of parameters and the stability of the topological phase is presented in the Supplemental Material \cite{supplement}.
\subsection{Second Chern number}
%
%
%
%
%
%
\begin{figure*}
	\includegraphics[width=0.8\textwidth]{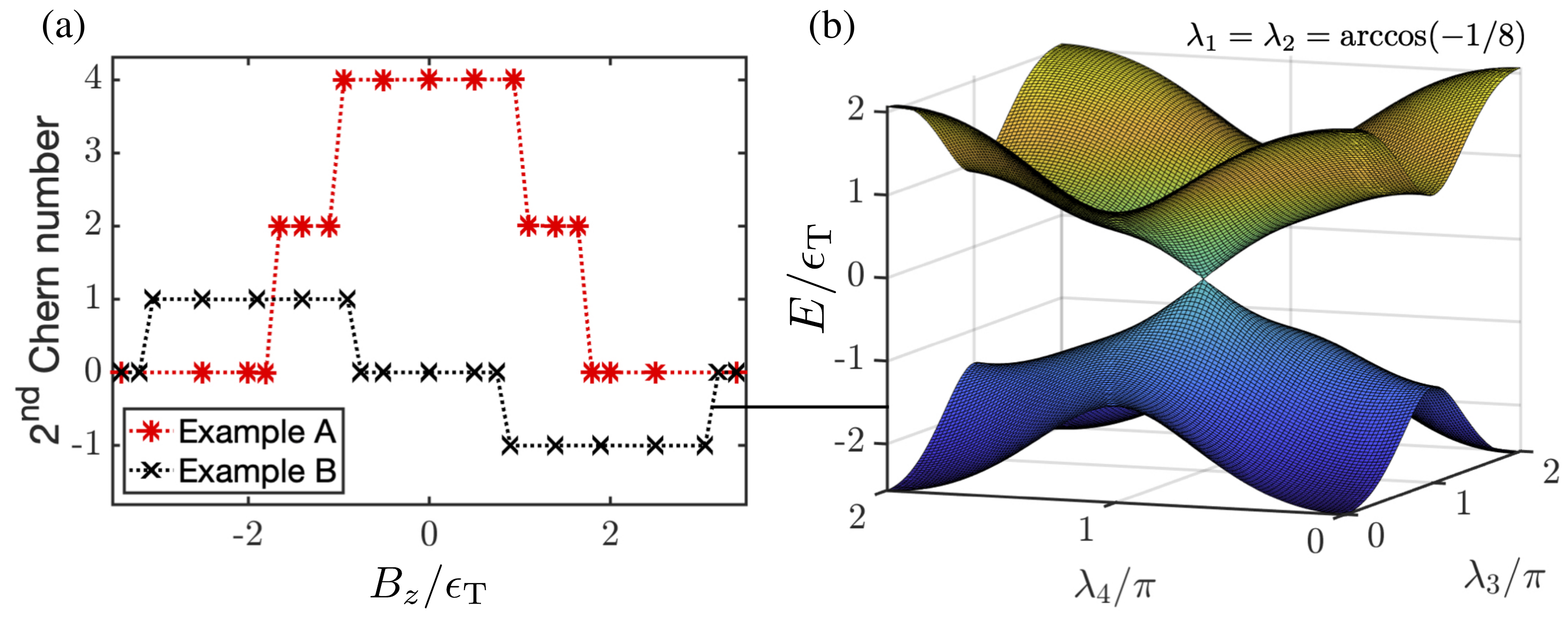}
	\caption{(a) Second Chern number numerically evaluated with Monte-Carlo integration for $N=10^9$ points in $\bm{\lambda}$ space. In Example A, we set $v_0=\epsilon_\mathrm{T}$, $\bm{w}_1/w_0 =(-1,0,1)$, $\bm{w}_2 / w_0 = (0,1,2)$ and $\bm{w}_3 / w_0 = (1,0,1)$.  In Example B, we set $\epsilon_\mathrm{T}=4\tilde{\epsilon}_\mathrm{T}/3 = \epsilon_\mathrm{h}$ and $h= \Delta/2$. (b) Energy bands with a crossing point for Example B at $B_z \approx 3.14\epsilon_\mathrm{T}$, which changes the second Chern number from $-1$ to $0$. In both cases $\epsilon_\mathrm{L} = \epsilon_\mathrm{R} = 0$.}
	\label{fig:Figure4}
\end{figure*}
%
%
%
%

%
The topological information of the system is encoded in the non-Abelian Berry curvature \cite{berry1984quantal,xiao2010berry} 
defined as \ $F_{jk} = \partial_j A_k - \partial_k A_j - i [A_j , A_k]$, where $A_j$ is the matrix valued non-Abelian Berry connection which has the elements $[A_j]_{\alpha\beta}=i\braket{\psi_{\alpha} | \partial_j \psi_\beta}$ calculated from the eigenstates $\ket{\psi_\alpha}$ of the degenerate energy levels and $\partial_j = \partial/\partial \lambda_j$. 
The first Chern number takes the usual form of $C^{(1_{jk})} = (1/2\pi) \int_{\mathbb{T}^2} \mathrm{d}^2\lambda \,\mathrm{Tr}(F_{jk})$ when integrating the Berry curvature over the 2-torus $\mathbb{T}^2$ spanned by the compact  parameters $(\lambda_j,\lambda_k)$, 
and the trace is taken with respect to the degenerate basis states.
However, it turns out that in our model with a pair of two degenerate levels, 
the first Chern number always vanishes \cite{sugawa2018second,manton2004topological}, $C^{(1_{jk})} = 0$.

Remarkably, the system is still topological in the sense of a higher-dimensional invariant.
In the four-dimensional space, we consider 
the second Chern number \cite{qi2008topological,avron1988topological}
%
%
%
%
%
%
%
%
\begin{align}
	C^{(2)} &= \frac{1}{32\pi^2} \int_{\mathbb{T}^4} \mathrm{d}^4 \lambda \, \varepsilon^{jklm} \,  \mathrm{Tr}\left(F_{jk}F_{lm}\right) ,
	\label{eq:2chern}
\end{align}
%
%
%
%
%
%
%
with the integration over the 4-torus $\mathbb{T}^4$ defined by the compact parameters $(\lambda_1,\lambda_2.\lambda_3,\lambda_4)$. Here, $\varepsilon^{jklm}$ is the Levi-Civita symbol in 4D and the sum runs over repeated indices.
It is worth to say that, in contrast to the first Chern number which is solely determined by the diagonal components of the Berry curvature, we need the additional information of the off-diagonal elements for the second Chern number.

In the following, we will show that the second Chern number does not vanish, $C^{(2)} \neq 0$, for some range of the parameters of the system.
An advantageous property  exists for  the symmetric case $\epsilon_\mathrm{L} = \epsilon_\mathrm{R} = \epsilon_0$.
Under this condition, the effective Hamiltonian in Eq.~\eqref{Hamstructure} can also be expressed by means of a set of five anti-commuting Dirac matrices 
$\bm{\Gamma} = (\Gamma_1, \ldots, \Gamma_5)$ with which we can write
$H = \bm{d}^\dagger\left(\epsilon_0 \openone + \bm{\Gamma} \cdot \bm{H}(\bm{\lambda}) \right)\bm{d}$, where the spinor of the dot operators reads
$\bm{d}^\dagger = (d_{\mathrm{L}\uparrow}^\dagger, d_{\mathrm{L}\downarrow}^\dagger, d_{\mathrm{R}\uparrow}^\dagger, d_{\mathrm{R}\downarrow}^\dagger)$. 
Here, $\openone$ is a $4\times 4$ unit matrix and the Hamiltonian describes a pseudospin $\bm{\Gamma}$ in an effective magnetic field  
$\bm{H}(\bm{\lambda}) = (H_1, H_2, H_3, H_4, H_5)$, where $H_{1}=B_z+Z$, 
$H_{2}=\mathrm{Re}[K]$, 
$H_{3}=\mathrm{Im}[K]$, 
$H_{4}=\mathrm{Re}[F]$ and 
$H_{5}=\mathrm{Im}[F]$.
Using this representation, 
the second Chern number can be written as a winding number of the 4D hypersurface traced out by $\bm{H}(\bm{\lambda})$ about the origin, i.e.,
%
%
%
%
%
%
%
%
\begin{align}
\label{eq:C_2_second}
C^{(2)}&=\frac{3}{8\pi^2}\int_{\mathbb{T}^4} \mathrm{d}^4\lambda \, \varepsilon^{jklmn} \frac{H_{j} (\partial_1 H_{k}) (\partial_2 H_{l}) (\partial_3 H_{m}) (\partial_4 H_{n}) }{|\bm{H}|^5} ,
\end{align}
%
%
%
%
%
%
%
counting how often the map $\bm{n} = \bm{H}/|\bm{H}|$ wraps around the unit sphere \cite{qi2008topological}.
Here, $\varepsilon^{jklmn}$ is the Levi-Civita symbol in 5D and the sum runs over repeated indices.
Using the expression in Eq.~(\ref{eq:C_2_second}), 
we numerically calculated the second Chern number by Monte-Carlo method \cite{montpack}
and report the result 
in Fig.~\ref{fig:Figure4}(a) for the specific examples A and B as a function of the externally applied magnetic field $B_z$.
We see that there are different topological phases with nonzero values of the second Chern number for both systems.
Interestingly, Example A admits a nontrivial second Chern number even in the absence of the external magnetic field $B_z=0$.
It is worth to analyze the behavior of the two energies associated with the twofold degenerate states, namely $E_{-}$ for the ground states and $E_{+}$ for the excited states. 
These two energies are two effective energy bands in a 4D space with the compact parameters $\lambda_1, \lambda_2, \lambda_3, \lambda_4$ playing the role of generalized quasimomenta in a first Brillouin zone.
Similarly to the behavior occurring in 2D topolgical systems characterized by the first Chern number, the topological transition with the concomitant discontinuous change of the second Chern number is signaled by the occurrence of crossing points between the two isolated bands $E_{-}$ and $E_{+}$ in the 4D space. 
In our effective model Hamiltonian describing the double-dot system, 
the upper band $E_{+}$ and the lower band $E_{-}$  have crossing points only if $|\bm{H}|=0$.
This equation defines the critical magnetic field $B_{z,c}$ as well as the critical values for the parameters $\bm{\lambda}_c = (\lambda_{1,c}, \lambda_{2,c}, \lambda_{3,c}, \lambda_{4,c} )$.
In order to illustrate this issue, we show the energies 
$E_{+}$ and $E_{-}$ in Fig.~\ref{fig:Figure4}(b) for Example B as a function of the two parameters $\lambda_3, \lambda_4$ at a fixed value of the other two parameters $\lambda_{1,c} = \lambda_{2,c} = \arccos(-1/8)$.
Notice that the two bands cross at the point $\lambda_{3,c} = \lambda_{4,c} = \pi $ with non-vanishing first derivatives.
Furthermore, the crossing does not happen at zero energy since the energy levels are renormalized in Example B.

\subsection{Non-Abelian Berry rotations in a degenerate subspace}
\label{nonA}
A cyclic change of the parameters $\bm{\lambda}(t)$ over time $t$ defining a periodic path in  parameter space is adiabatically slow if the characteristic scale variations of the  parameters $\partial \lambda_j/\partial t$ $(j=1,\dots,4)$ remain much smaller than the energy gap $(E_{+}-E_{-}) \gg \hbar|\partial \lambda_j/\partial t|$ between the first excited state with energy $E_{+}$ and the ground state with energy $E_{-}$ (with $\hbar$ being Planck's constant). 
If the ground state is nondegenerate, the final state acquires a complex geometric phase, known as the Berry phase \cite{berry1984quantal} at the end of the cyclic path.
For a degenerate ground state, the non-Abelian Berry phase due to a cyclic path represents a unitary transformation of the initial state in the degenerate subspace \cite{wilczek1984app,pachos1999non}.
In our system, the ground state is spanned by two eigenstates and the cyclic path results in a rotation in this two-dimensional degenerate subspace.
The adiabatic condition guarantees that the final state has vanishing components in the twofold degenerate excited state after the cyclic path, namely Landau-Zener transitions are negligible.

Consider the closed path $\mathrm{C}_p: t \mapsto \boldsymbol{\lambda}^{(p)}(t)$ for the time range 
$t \in [0,T]$ with the periodic condition $\bm{\lambda}^{(p)}(0) = \bm{\lambda}^{(p)}(T)$. 
Defining a specific basis in the ground state subspace $\ket{E_-^{(\alpha)}}$ ($\alpha = 1,2$) and assuming that the initial ground state is $\ket{\Psi_\mathrm{i}}=\ket{E_-^{(1)}}$, the final ground state $\ket{\Psi_\mathrm{f}}$ after the evolution becomes the superposition
%

%
%
%
%
\begin{figure*}
	\includegraphics[width=0.8\textwidth]{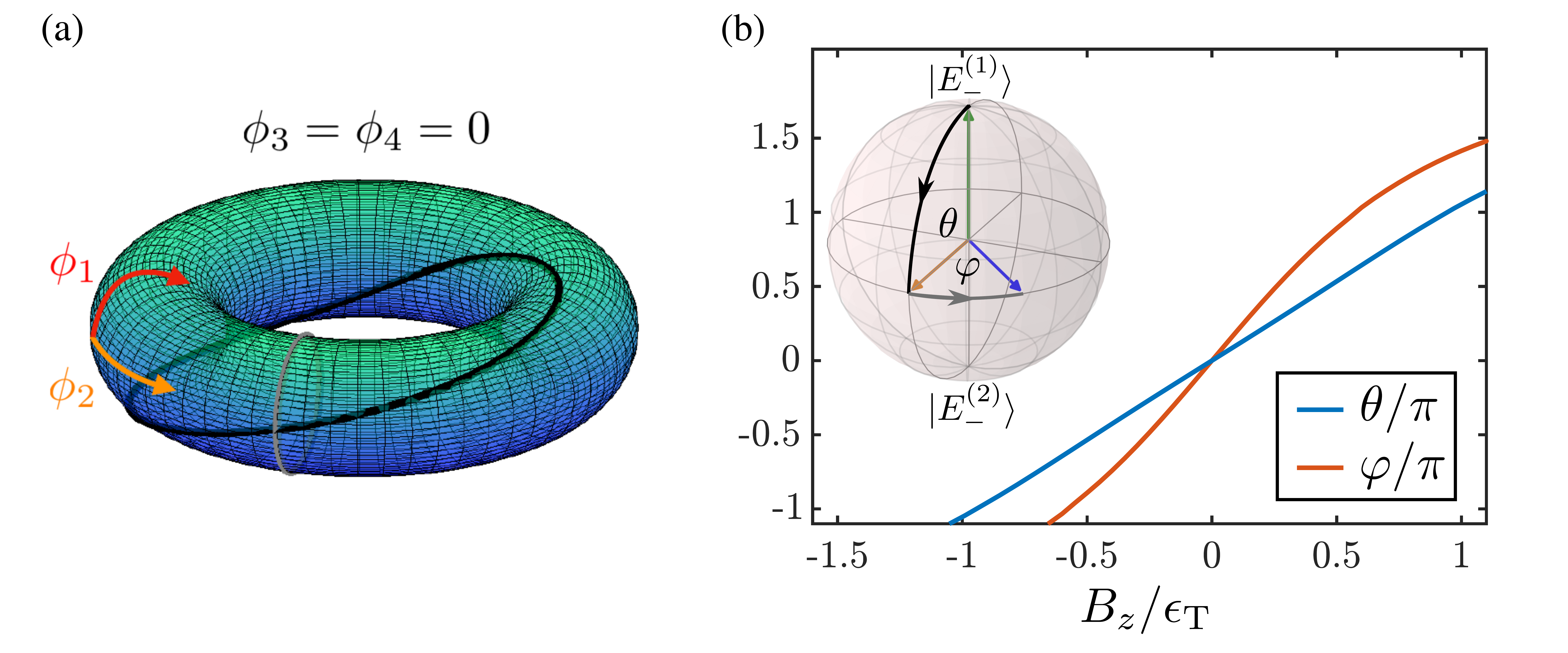}
	\caption{(a)
		Protocol for qubit manipulations for any initial state (here $\ket{E_-^{(1)}}$) via the non-Abelian Berry phase in the case of Example A. The rotation from $\ket{E_-^{(1)}}$ to $\ket{E_-^{(2)}}$ (black) for an adiabatic change of $\phi_1,\phi_2:0\rightarrow 2\pi$ is depicted on a torus. A second rotation around the $z$-axis of the Bloch sphere (grey) can be achieved via the adiabatic change of $\phi_1:0\rightarrow 2\pi$ and $\phi_2=0$. $\phi_3=\phi_4=0$ during the whole process. (b) Bloch sphere of the degenerate subspace $E_-$ with the possible rotations in (a) (top left corner). Numerically determined angles of the Berry rotations $\varphi$ (by changing adiabatically $\phi_1,\phi_2:0\rightarrow 2\pi$) and $\theta$  (by changing adiabatically $\phi_1:0\rightarrow 2\pi$) as a function of the magnetic field $B_z$ in the case of Example A for $v_0 = \epsilon_\mathrm{T}$, $\bm{w}_1 / w_0 = (-1,0,1)$, $\bm{w}_2 / w_0 = (0,1,2)$, $\bm{w}_3 / w_0 = (1,0,1)$, and $\epsilon_\mathrm{R}=\epsilon_\mathrm{L}=0$.}	
	\label{fig:Figure5}
\end{figure*}
%
%
%
%
%
%
%
%
%
%
%
%
\begin{align}
\ket{\Psi_\mathrm{f}} = \exp\Bigl( -\frac{i}{\hbar} \int_0^T \mathrm{d}t' \, E_-(\bm{\lambda}^{(p)}(t')) \Bigr)
\left(\gamma_{11} \ket{E_-^{(1)}}+\gamma_{21}\ket{E_-^{(2)}}\right)\, ,
\end{align}
%
%
%
%
%
%
%
in which the first term is the (scalar) dynamical phase and the coefficients of the linear combination are given by the matrix elements of the unitary rotation as
%
%
%
%
%
%
%
%
\begin{align}
\gamma_{\alpha\beta}= \left[\mathcal{P} \exp\left(\oint_{\mathrm{C}_p} \mathrm{d}\bm{\lambda} \cdot \bm{A}_{-}(\bm{\lambda})\right)\right]_{\alpha\beta} \, ,
\end{align}
namely the non-Abelian Berry phase describing the rotation.
Here, $\bm{A}_{-}$ is the non-Abelian Berry connection in the degenerate ground-state subspace and $\mathcal{P}$ is the path-ordering operator.
In general, due to the rotational character, the local and infinitesimal  non-Abelian Berry rotations do not commute along the path $\mathrm{C}_p$ which makes it necessary to use the path-ordering operator $\mathcal{P}$.

The preparation of the initial state can be  simply achieved in the following way.
Initially, one can introduce a small asymmetry  $\epsilon_\mathrm{L} \neq \epsilon_\mathrm{R}$ in the two dots  such that 
it lifts the degeneracy of the energy levels of the ground state in the symmetric case.
This process unequivocally identifies the nondegenerate states $\ket{E_-^{(1)}}$ and $\ket{E_-^{(2)}}$ for    $\epsilon_\mathrm{L} \neq \epsilon_\mathrm{R}$. 
After waiting a sufficiently long time much larger than the time scale set 
by relaxation processes, one can ensure that state is in the lowest state which we label $\ket{E_-^{(1)}}$. 
Subsequently, the degeneracy is adiabatically restored by tuning $\epsilon_\mathrm{L} = \epsilon_\mathrm{R}$, such that the system remains in the $\ket{E_-^{(1)}}$ state.
Then, one can make use of the non-Abelian Berry phase to create arbitrary rotations in the degenerate subspace by simply adiabatically changing  the parameters along a closed loop. 
To give an idea of such implementations, we discuss an example of a state-preparation protocol in detail for the specific case corresponding to the system A, see Fig.~\ref{fig:Figure3}(a).
In Fig.~\ref{fig:Figure5}(a), we show two adiabatic cyclic paths in which two SC phases are always fixed, $\phi_3 = \phi_4 = 0$, and we adiabatically change the other two phases $\phi_1$ and $ \phi_2$.
The first path is defined by the simultaneous change of the two phases with $\phi_1 = \phi_2 = \phi$ along the path $\phi =  0 \rightarrow 2\pi$, depicted in  Fig.~\ref{fig:Figure5}(a) as a black line.
For the second path, as presented in Fig.~\ref{fig:Figure5}(a) by a grey line, we only vary the phase $\phi_1$ and fix the other phase $\phi_2 = 0$.
Representing the degenerate ground state on a Bloch sphere with the basis states $\{ \ket{E_-^{(1)}} , \ket{E_-^{(2)}} \}$, these two adiabatic cyclic paths implement qubit rotations, as presented in Fig.~\ref{fig:Figure5}(b). 
While the first black path results in a rotation described by the polar angle $\theta$, the second grey path yields a rotation in the azimuthal direction described by the angle $\varphi$, where both values of the angles $\theta$ and $\varphi$ depend on the value of the externally applied magnetic field $B_z$ in the two dots, as shown in Fig.~\ref{fig:Figure5}(b).
A desired combination of the two paths, achieved by previously fixing the externally applied magnetic field to the desired value, one can create arbitrary rotations in the degenerate subspace.\\
Similar results hold for Example B, see Fig.~\ref{fig:Figure3}(b).
In this case one can also vary the angles of the local magnetization in the hybrid superconductor-ferromagnet contacts, $\theta_1$ and $\theta_2$.
For instance, the polar rotation is set by adiabatically varying $\theta_1$ and $\theta_2$ along the cyclic path defined by $\theta_1 = \theta_2 = 0\rightarrow 2\pi$ and by setting the SC phases to $\phi_1=\phi_2=0$, whereas the azimuthal rotation is implemented by varying only $\phi_2 = 0 \rightarrow 2\pi$ and by simply setting $\phi_1 = \theta_1 = \theta_2 =0$.
Clearly, the angle-dependence on the local magnetic field $B_z$ is different in the case of Example B in comparison to Example A and depends on the concrete parameters of the setup. The rotation angles of Example B are further discussed in the Supplemental Material \cite{supplement}.
%
%
%
%
\begin{figure*}
	\includegraphics[width=0.9\textwidth]{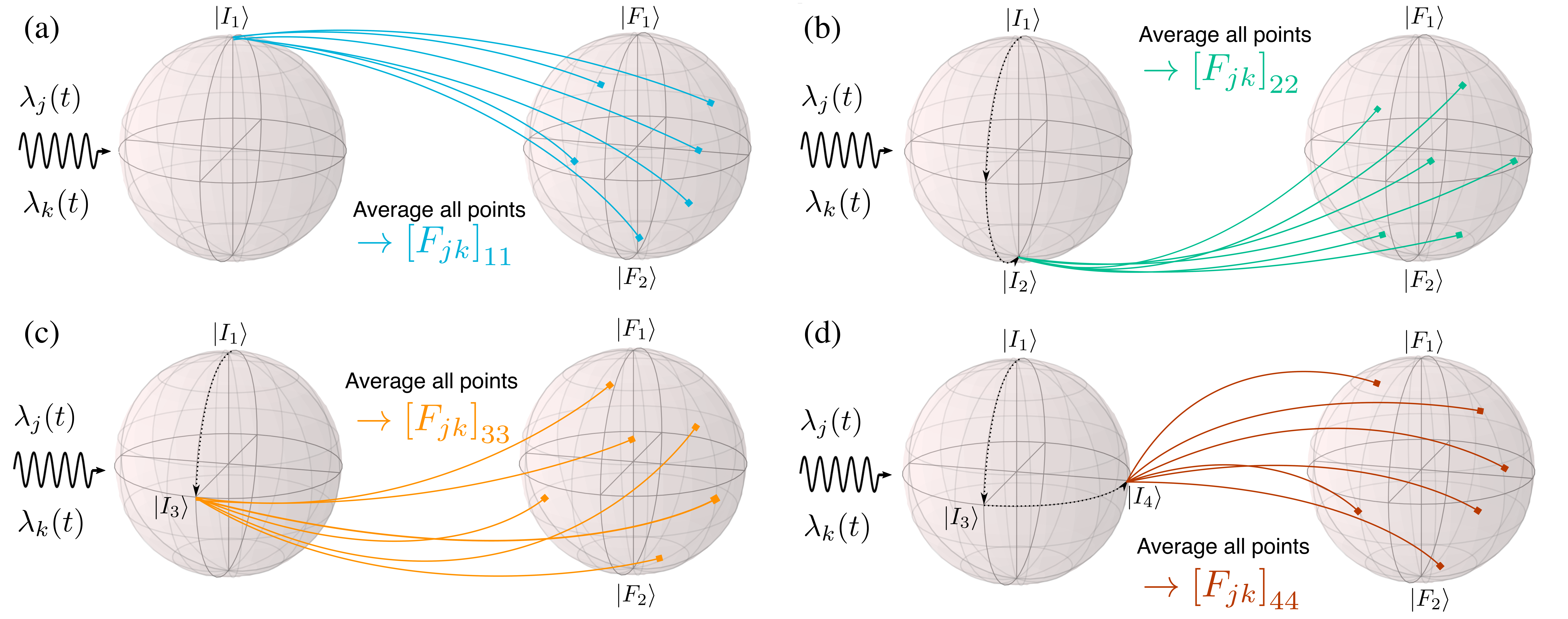}
	\caption{(a), (b) Microwave spectroscopy protocol for the diagonal elements of the non-Abelian Berry curvature in the degenerate subspace. A small time dependent modulation is applied to two parameters $\lambda_j(t) = 2A\cos(\omega t)/\hbar\omega$ and $\lambda_k(t) = 2A\cos(\omega t\pm\pi/2)/\hbar \omega$, $j\neq k$, with $A/\hbar\omega\ll 1$, leading to a transition from the initial state $\ket{I_j}$ to an unknown final state $\ket{F}$. From the averaged response over the excited states the diagonal part of the Berry curvature can be measured. Similarly, the diagonal Berry curvature with respect to the rotated states $\ket{I_3}$ (c) and $\ket{I_4}$ (d) can be measured. From these, the off-diagonal part of the non-Abelian Berry curvature with respect to the initial basis can be constructed by means of Eq. \eqref{eq:offdiagonal}.}
	\label{fig:Figure6}
\end{figure*}
%
%
%
%
%
%
%
\subsection{Measuring the second Chern number using polarized microwave spectroscopy}

We now discuss how to experimentally measure the second Chern number in higher-dimensional topological SC systems.
The measurement scheme is based on microwave spectroscopy in which designed microwave excitations are applied to the system. 
In particular, we show how to get access to the local elements of the non-Abelian Berry curvature by means of polarized microwave spectroscopy from which the second Chern number automatically follows. 
This task corresponds to a nontrivial extension of the method proposed 
in SC multiterminal Josephson junctions with lower dimensional topology, i.e., nonvanishing first Chern number, with the ground and the excited states being nondegenerate \cite{klees2020microwave}.

Applying a small time-dependent periodic modulation of driving frequency $\omega$ to the two parameters $\lambda_j$ and $\lambda_k$ of the Hamiltonian in Eq.~\eqref{Hamstructure} and with a fixed phase difference $\delta$, we have
%
%
%
%
%
%
%
%
\begin{align}
H \to H + \frac{2A}{\hbar \omega} \frac{\partial H}{\partial \lambda_j} \cos(\omega t) + \frac{2A}{\hbar \omega} \frac{\partial H}{\partial \lambda_k} \cos(\omega t - \delta) ,
\label{eq:microwaveModulation}
\end{align}
with the condition of low microwave power $A \ll \hbar \omega$.
In general, discrete lines appear in the microwave absorption spectrum originating from the transitions from the ground-state subspace to the subspace of excited states.
For a weak perturbation, the intensity of these lines is quantified by the transition rates which can be computed using the Fermi's Golden Rule, following  standard linear response theory.
However, as the ground-state subspace is twofold degenerate, one can prepare different combinations of the initial state before applying microwave radiation, as illustrated in Fig.~\ref{fig:Figure6}.
Therefore, we have to deal with different transition rates depending on the specific initial state. 
Instead of measuring the absorption spectrum, the transition rates for each initial state have to be extracted by analyzing the switching statistics of the single-photon absorption event \cite{Hays:2019} and by averaging over the final state.

As discussed above, starting from the asymmetric case $\epsilon_\mathrm{L} \neq \epsilon_\mathrm{R}$ one can univocally define the basis state in the ground and excited subspaces which we denote $\ket{E_{\pm}^{(\alpha)}}$ $(\alpha = 1,2)$, see Fig.~\ref{fig:Figure6}.
Then, in order to measure the off-diagonal elements of the non-Abelian Berry curvature, we need to perform four different absorption measurements in which, for each of them, we have to initialize the system in the following four ground states: $\ket{I_1} = \ket{E_-^{(1)}}$ in  Fig.~\ref{fig:Figure6}(a), $\ket{I_2} = \ket{E_-^{(2)}}$ in Fig.~\ref{fig:Figure6}(b), $\ket{I_3} = (\ket{E_-^{(1)}} + \ket{E_-^{(2)}})/\sqrt{2}$  in Fig.~\ref{fig:Figure6}(c), and $\ket{I_4} = (\ket{E_-^{(1)}} + i \ket{E_-^{(2)}})/\sqrt{2}$ in Fig.~\ref{fig:Figure6}(d).
This is the minimal amount of measurements required to extract the non-Abelian Berry curvature \cite{kolodrubetz2016measuring}.
Each state $\ket{I_{\mu}}$ ($\mu=1,2,3,4$) can be prepared using the previously discussed rotation protocols which are achieved by tailored adiabatic cyclic paths of the controlling parameters.

In general, applying the microwave signal at two parameters $\lambda_j$ and $\lambda_k$, as described in Eq.~\eqref{eq:microwaveModulation}, results in a transition from the prepared initial state $\ket{I_{\mu}}$ to an arbitrary combination of the excited states, namely $\ket{F} = \cos\left(\theta_{F}/2\right) \ket{E_+^{(1)}} + \sin\left(\theta_{F}/2\right) \, e^{i \varphi_{F}} \ket{E_+^{(2)}}$, as shown in Fig.~\ref{fig:Figure6}, with the angles $\theta_{F}$ and $\varphi_{F}$ on the Bloch sphere of the excited subspace.
The transition rates associated with the switching of the states are proportional to \cite{klees2020microwave,Ozawa:2018} 
%
%
%
%
%
%
%
\begin{align}
\label{eq:Mjk}
M_{jk,\mu}^{(\delta)} (\theta_F , \varphi_{F})
&= 
\braket{I_\mu| \partial_j H | F }  \braket{ F | \partial_j H  | I_\mu} 
\nonumber \\
&\quad + \braket{I_\mu| \partial_k H | F }    \braket{F |  \partial_k H | I_\mu} 
\nonumber \\
&\quad + e^{i \delta} \braket{I_\mu| \partial_j H | F }  \braket{F |  \partial_k H | I_\mu} 
\nonumber \\
&\quad + e^{-i \delta} \braket{I_\mu| \partial_k H | F }  \braket{F | \partial_j H  | I_\mu} .
\end{align}
%
%
%
%
%
%
One can measure the delay times from an application of the microwave drives until the system gets excited.
By repeating this protocol several times and by analyzing the statistical distribution of these switching events, one can extract the average transition rates which are proportional to the matrix elements of Eq.~\eqref{eq:Mjk}.
Finally, since we do not exactly know the final state, we consider the average of such matrix elements over all possible final states, namely over $\theta_{F}$ and $\varphi_{F}$.
A priori, this should correspond to a real implementation of the measurement in which one repeats the single-photon absorption event several times.  

As the matrix elements depend on the relative phase lag $\delta$ between the two modulations, one can choose $\delta = + \pi /2$  and  $\delta = - \pi /2$ to implement circularly polarized microwaves.
The difference between these averaged matrix elements is related to the diagonal component of the local non-Abelian Berry curvature for the particular initial state $\ket{I_\mu}$, i.e.,
%
%
%
%
%
%
%
\begin{widetext}
\begin{align}
\label{eq:difference}
\frac{1 }{ 4\pi }& \int_0^\pi \!\!\! \mathrm{d}\theta_F \int_0^{2\pi} \!\!\! \mathrm{d}\varphi_F  \,  \sin\theta_F \, 
\Bigl[ 
M_{jk,\mu}^{(\pi/2)}(\theta_F , \varphi_{F}) 
- M_{jk,\mu}^{(-\pi/2)}(\theta_F , \varphi_{F})
\Bigr] 
= 4E_-^2 \, \left[F_{jk}\right]_{\mu \mu}.
\end{align}
\end{widetext}
In this way, one has access to the diagonal components of the non-Abelian Berry curvature in the  ground-state subspace with basis of the initial states with $\mu=1$ or $\mu=2$.
The off-diagonal elements of the non-Abelian Berry curvature $[F_{jk}]_{12}$ and $[F_{jk}]_{21}$ in this basis can be obtained using the following properties \cite{kolodrubetz2016measuring}
%
%
%
%
%
%
%
\begin{subequations}
	\label{eq:offdiagonal}
	\begin{align}
	\left[F_{jk} \right]_{12}&= \left[F_{jk} \right]_{33}  -  i \left[F_{jk} \right]_{44}-  \frac{(1- i)}{2} \left(\left[F_{jk} \right]_{11} + \left[F_{jk} \right]_{22} \right) ,
	\\
	\left[F_{jk} \right]_{21}&= \left[F_{jk} \right]_{33} + i \left[F_{jk} \right]_{44} - \frac{(1+ i)}{2} \left(\left[F_{jk} \right]_{11}+ \left[F_{jk} \right]_{22} \right)  \, ,
	\end{align}
\end{subequations}
%
%
%
%
%
%
%
in which $[F_{jk}]_{33}$ and $[F_{jk}]_{44} $ are given by Eq.~(\ref{eq:difference}) with the initial states $\mu=3$ and $\mu=4$.
In simple words, one has to repeat the same measurement for all four different states in order to reconstruct all the elements of the non-Abelian Berry curvature matrix for a given modulation of $\lambda_j$ and $\lambda_k$.

In summary, one has to prepare the system in a state $\ket{I_\mu}$ with the help of the adiabatic cyclic path protocol presented in the previous section and apply circular microwave drives to $\lambda_j$ and $\lambda_k$ with $\delta = \pm \pi / 2$.
The transition rates for the single absorption event are extracted by studying the switching statistics.
By taking the difference between transition rates of left ($+\pi/2$) and right ($-\pi/2$) circularly polarized microwaves on the initial states for $\mu=1,2$, we obtain the diagonal elements of the Berry curvature $[F_{jk}]_{11}$ and $[F_{jk}]_{22}$, whereas the rate associated to the initial states $\mu=3,4$ allows us to reconstruct the off-diagonal elements $[F_{jk}]_{12}$ and $[F_{jk}]_{21}$ according to Eq.~\eqref{eq:offdiagonal}.
This has to be repeated for all $j,k=1,2,3,4$ with $j<k$ to find all Berry curvatures of the 4D parameter space $\bm{\lambda}$, from which the second Chern number follows by integration according to Eq.~\eqref{eq:2chern}.

As a final remark, notice that if just a single parameter $\lambda_j$ is modulated or $\delta = 0$ and $\delta = \pi$ are chosen, the ground state metric tensor $g_{jk}$ can be measured aswell, see Refs.~[\onlinecite{klees2020microwave}] and [\onlinecite{Ozawa:2018}].
\section{Discussion}
We have presented a first theoretical proposal for a nontrivial second Chern number in SC quantum systems, realizable via a double-dot system and an environmental network structure formed by SC contacts.
The need to operate in the odd-parity regime with a large Coulomb interaction on the dots to get rid of Cooper pair injections can be also advantageous, as it possibly restricts dissipation from quasiparticle excitations by restricting parity changes in the system with the inter- and intra-dot Coulomb interaction.

The parameter space is not necessarily restricted to SC phases only, as we have shown in Example B in which a part of the space is analogously created by the magnetization angles of the leads.
This yields a great freedom for eventually finding much simpler environmental networks with the presented structure in Eq.~\eqref{Hamstructure}, where similar protocols for the non-Abelian Berry phase and the measurement of the second Chern number should hold, as they can be generally applied on such parameters. 
Furthermore, the non-Abelian Berry phase is potentially useful to create arbitrary rotations in a given subspace of degenerate quantum states. 
Ultimately, this could be used for holonomic quantum computation to create a universal set of quantum gates for a new type of topological qubit with an emergent second Chern number. 
Hence, a further direction of interest could be the stability of the non-Abelian Berry phase under local fluctuations of the system, which would define how stable and topologically protected these protocols for the quantum gates would be in non-ideal systems.
To conclude, we emphasize that this could stimulate further research activities for the investigation of higher-dimensional topology and new types of topological phases in Josephson matter, which seems easily accessible in such systems due to the, in general, unlimited dimensionality.
\begin{acknowledgements}
The authors are grateful for funding provided by the DFG through SFB 767 and Grant No. RA 2810/1.\\
\end{acknowledgements}

\end{document}